\documentstyle[12pt]{article}
\newcommand{\bea}{\begin{equation}}
\newcommand{\eea}{\end{equation}}
\newcommand{\ber}{\begin{eqnarray}}
\newcommand{\eer}{\end{eqnarray}}
\begin{document}
\title{\bf Wall Bounded Turbulent Shear Flow: Analytic Result For An Universal Amplitude}
\author{Amit Kr. Chattopadhyay, Jayanta K. Bhattacharjee \\
Dept. Of Theoretical Physics,\\
Indian Association For The Cultivation Of Science,\\
Jadavpur, Calcutta 700 032, India}
\date{\today}
\maketitle
\begin{abstract}
In the turbulent boundary layer above a flat plate, the velocity profile is
known to have the form $ v=v_0 [\frac{1}{\kappa}ln z+constant] $. The
distance from the wall in dimensionless units is $ z $ and $ v_0 $ is an
uniquely defined velocity scale. The number $ \kappa $ is universal and
measurements over several decades have shown that it is nearly 0.42. We use
a randomly stirred model of turbulence to derive the above law and find
$ \kappa=\sqrt{\frac{108}{125 \pi}} \simeq 0.52 $.\\\\
PACS number(s): 47.27.-i, 47.27.Nz, 64.60.Ht
\end{abstract}
\newpage
The turbulent flow over a stationary plate has been investigated in great detail for well over a century. One of the important quantities studied is the velocity profile and in particular a good deal of attention has been paid to the velocity in the boundary layer. The existence of the boundary layer was first postulated by Prandtl [1,2]. Due to the presence of  viscosity, the flow speed has to be zero on the surface of the plate. Within a short distance, the velocity has to rise to its bulk value. This small distance in which the velocity experiences a rapid change is called the boundary layer. Prandtl observed that in the boundary layer the velocity $ v $ scales with the distance $ z $ from the boundary. Adjoining the boundary is an extremely thin laminar boundary layer, after which comes a buffer zone and then the turbulent boundary layer. In this regime one has  [2,3]
\bea
\frac{v}{v_0}\:=\:\frac{1}{\kappa}[ln(z/z_0)] ,
\eea 
a result known as the law of the wall, where $ v_0 $ is a reference velocity to be defined below and $ \kappa $ is an universal constant which so far has only been experimentally measured [4].
\par
With the coming of Kolmogorov's theory [5] of homogeneous isotropic turbulence, where simple dimensional argument yielded remarkable results, dimensional arguments were similarly used in this problem to obtain eqn.(1). A more careful analysis [6,7] of the Navier Stokes equation also supported the above result, which is one of the cornerstones in the theory of the turbulent boundary layer. Over the last two decades, a lot of effort has been spent in providing a more microscopic justification for Kolmogorov's scaling in homogeneous isotropic turbulence, thereby enhancing our understanding of the energy cascade involved. Analysis of homogeneous isotropic turbulence has thus become a problem of non-equilibrium statistical physics [8-14]. In this work, we propose to use techniques from statistical physics to study the law of the wall and in the process provide calculations for the universal constant $ \kappa $.
\par
We begin by explaining the geometry and the traditional formulation. The flat plate (infinitely large) sits at $ z=0 $ (in the x-y plane) and the flow (in the steady laminar situation) occurs in the x-direction, due to some maintained pressure gradient in that direction. We can view this as a two dimensional problem -- the dynamics occurs only in the x-z plane. Continuity (for an incompressible flow) implies $ \frac{\partial u}{\partial x}\:+\:\frac{\partial w}{\partial z}\:=\:0 $. Translational invariance in the x-direction makes quantities x-independent and hence $ \frac{\partial w}{\partial z}\:=\:0 $. The resulting constant value of $ w $ is found to be zero since $ w\:=\:0 $ at $ z\:=\:0 $. In the laminar regime (low Reynold's number) the velocity $ u $ satisfies $ \bigtriangledown^2 u\:=\:0 $ and since the only dependence of $ u $ is on $ z $, we have $ \frac{\partial^2 u}{\partial z^2}\:=\:0 $ or a linear profile for $ u $, i.e. $ u\:=\:u_0\frac{z}{a} $, where $ u_0 $ and $ a $ are constants. This is the usual Couette flow.
\par
We now pass to high Reynold's number where the non-linear terms in the Navier Stokes' equation need to be retained. We assume that the flow is turbulent, so that it pays to write the velocity field as $ \vec v=\vec V+\vec u $, where $ <\vec v>\:=\:\vec V $. This has the implication that $ \vec u $ (components $ u_x, u_y, u_z $; which from now onwards we will designate as $ u_1, u_2, u_3 $ respectively) is a purely fluctuating component and $ \vec V $ (components $ V_x $ and $ V_z $, henceforth to be called as $ V_1, V_3 $ respectively) is the mean flow. No mean quantity will depend on $ x $ due to translational invariance and hence $ V_z=0 $ i.e. $ V_3=0 $, by the same argument as given above. Hence finding the flow profile is the same as finding $ V_1(x_3) $ (i.e. $ V_x(z) $).
\par
Writing the Navier Stokes' equation as
\bea
\frac{\partial \vec v}{\partial t}\:+\:(\vec v.\vec\nabla)\vec v\:=\:-\frac{{\vec\nabla} P}{\rho}\:+\:\nu \bigtriangledown^2 \vec v,
\eea
substituting $ \vec v\:=\:\vec V\:+\:\vec u $ and taking an average yields (for very high Reynold's number) after standard manipulations [7]
\bea
\nu \frac{\partial V_1}{\partial z}\:=\:<u_1 u_3>\:+\:constant
\eea
At this point, one invokes dimensional arguments to argue $ V_1\:=\:\frac{V_0}{\kappa}ln(z/z_0) $. Our goal is to calculate $ <u_1 u_3> $ directly.
\par
We proceed by writing an equation of motion for the fluctuating velocity field $ \vec u $. This is the usual Navier Stokes' equation supplemented by a random forcing term which is generated by the interaction of the mean  flow and the turbulent flow and the boundary conditions. We write this as
\bea
\frac{\partial u_i}{\partial t}\:+\:u_j \frac{\partial}{\partial x_j} u_i\:=\:-\frac{1}{\rho}\frac{\partial P}{\partial x_i}\:+\:\nu \bigtriangledown^2 u_i\:+\:f_i
\eea
with $ \frac{\partial u_i}{\partial x_i}=0 $ (this determines $ P $ in terms of $ u_i $ and makes $ \frac{\partial P}{\partial x_i} $ a non-linear term), where $ f_i $ is the random force, whose correlation is to be specified. For the Kolmogorov theory of homogeneous isotropic turbulence, this forcing gives rise to spatially correlated noise [9-14] and in a D-dimensional space, the Fourier components $ f_i(k,\omega) $ of $ f_i(x,t) $ satisfy $ <f_i(k_1,\omega_1)f_j(k_2,\omega_2)>\:=\:F(k_1)\delta(k_1+k_2)\delta(\omega_1+\omega_2)P_{ij}(k_1) $ where $ F(k)\:\sim\:k^{-D} $ and $ P_{ij}(k) $ is the projection operator. In co-ordinate space  $ <f_i(x_1,t_1) f_j(x_2,t_2)>\:\propto\:P_{ij}\tilde F(x_{12}) \delta (t_1-t_2) $ where $ \tilde F(x_{12}) $ is a constant for $ x_{12} \rightarrow 0 $. The combination of the non-linear terms and the forcing changes the basic relaxation rate in the problem. In momentum space, in the absence of the non-linear terms, the basic frequency is  clearly $ \nu k^2 $. The non-linear terms constitute a relevant perturbation and makes $ \nu $ diverge for small $ k $.For the forcing of $ k^{-D} $, one gets the Kolmogorov time scale where $ \nu\:=\:\overline{\nu_0} k^{-4/3} $ and the basic frequency is $ \overline{\nu_0} k^{2/3} $ when the sweeping effects are removed. For this problem of calculating $ <u_1 u_3> $, sweeping effects will set the time scale (interaction of large and small eddies) and hence the basic frequency scale is $ \nu_0 k $ [13,14]. It should be noted that for homogeneous isotropic turbulence the cross correlation $ <u_1(\vec r,t) u_3(\vec r,t)> $ will be zero -- the finite value of the correlation is because of the breaking of isotropy by the rigid boundary at $ z=0 $. 
\par
To deal with the situation, restricting ourselves to $ z \geq 0 $, we need to use field variables for which we introduce Fourier transforms in the x-y plane and keep the co-ordinate description along the z-axis. To take into account the solenoidal condition, it is convenient to work in the potential $ \vec A $, so that $ \vec u\:=\:\vec\nabla \times \vec A $.
We expand $ \vec A $ as
\bea
\vec A_i(\vec r,t)\:=\:\sum_{\vec k,\omega}\:\vec A_i(\vec k,z,\omega)e^{i(\vec k.\vec r-\omega t)}
\eea
and thereafter write down the equation of motion for $ \vec A $ from eqn.(4). We will denote the random force in the equation of motion for $ \vec A $ by $ \vec F $. Specifying the correlations of $ \vec F $ will now be the main task. One part of the correlations of $ F_i $ and $ F_j $ will be the homogeneous isotropic part, giving $ <F_i(k,z_1) F_j(- k,z_2)>\:=\:\delta_{ij} \frac{1}{k^4} F(z_{12}) $ where $ F(z_{12}) $ is a constant for the spatially correlated noise. This reproduces the Kolmogorov spectrum. The boundary at $ z=0 $, will cause anisotropy in the correlation of $ F_i $ in the x-y plane (this anisotropy causes the correlation $ <u_1 u_3> $ to be non-zero in the presence of the wall). Considering all these factors, the final correlations of $ \vec F $ assume the form
\bea
<F_i(k,z_1) F_j(k,z_2)>\:=\:A\frac{\epsilon}{k^4}\:[\delta_{ij}\:+\: \delta_{i1} \delta_{j3}\:+\:\delta_{i2} \delta_{j3}]
\eea
where $ \epsilon $ is the rate of production of turbulent energy. The first term on the right hand side produces the isotropic Kolmogorov response for the fluctuating field. The second and third terms reproduce the effect of the boundary.
\par
We now start with a linear equation of motion for $ A_i(k,z) $ with the effect of the non-linear terms included in an effective viscosity $ \nu_{eff} $. This reads
\bea
\frac{\partial A_i}{\partial t}\:+\:\nu_{eff}(k^2-\frac{\partial^2}{\partial z^2})A_i\:=\:F_i
\eea
with the correlations of $ F_i $ specified in eqn.(6). We see from eqn.(3) that our primary goal is to calculate the stress tensor $ \tau_{13}\:=\:<u_1 u_3> $. This can be written as
\ber
<u_1(\vec r,t) u_3(\vec r,t)>&=&\int\:\frac{d^2k}{(2\pi)^2} \frac{d\omega}{2\pi}\:<u_1(k,z,\omega) u_3(-k,z,-\omega)>
\nonumber \\
&=&-\int\:\frac{d^2k}{(2\pi)^2} \frac{d\omega}{2\pi}\:k_{2}^2\:<A_3(k,z,\omega) A_1(-k,z,-\omega)> \nonumber \\
&=&-A \epsilon\:\int\:\frac{d^2k}{(2\pi)^2} \frac{d\omega}{2\pi}\:\frac{k_{2}^2}{k^4}\:\int\:\frac{G_{+}(z,z_1) G_{-}(z,z_2)}{\nu_{eff}^2}dz_1 dz_2 \nonumber \\
\eer
where $ G_{\pm}(z,z\prime) $ are the Green's functions associated with the operator $ L\:=\:\frac{\partial^2}{\partial z^2}-k^2\:\mp\:i\Omega $, where $ \Omega=\omega/\nu $. In arriving at the last line of eqn.(8), we have used the correlation of eqn.(6) and noted the fact that the angular average of $ k_1 $ is zero. Defining $ \kappa_{\pm}^2\:=\:k^2 \pm i\Omega $ and noting that we are interested in a single rigid boundary, $ G_{\pm} $ can be written down as [17]
\bea
G_{\pm}(z,z\prime)\:=\:\frac{sinh(\kappa_{\pm}z_{<})\:e^{-\kappa_{\pm}z_{>}}}{\kappa_{\pm}}
\eea
Carrying out the integration in eqn.(8), we have 
\ber
<u_1 u_3>\:&=&\:-A \epsilon\:\int\:\frac{d^2k}{(2\pi)^2} \frac{d\omega}{2\pi}
\:\frac{k_2^2}{k^4}\:\frac{1}{\kappa_{+}^2 \kappa_{-}^2} \nonumber \\
& & \frac{1}{\nu_{eff}^2}\:(1-e^{-\kappa_{+}z})(1-e^{-\kappa_{-}z})
\eer
For large $ z $, the stress tensor tends to a constant value $ -V_0^2 $ which is given by
\ber
V_0^2\:&=&\:A \epsilon\:\int\:\frac{d^2k}{(2\pi)^2}\frac{d\omega}{2\pi}\:\frac{cos^2{\theta}}{k^2}\:\frac{1}{\nu_{eff}^2}\:\frac{1}{\kappa_{+}^2 \kappa_{-}^2}
\nonumber \\
&=&\frac{A\epsilon}{2}\:\int\:\frac{d^2k}{(2\pi)^2}\:\frac{cos^2{\theta}}{\nu_{eff}}\:\frac{1}{k^4}
\eer
[The integrals have to be cut-off at a low momentum $ k_0 $.]\\
In the sweeping time dominated situation $ \nu_{eff}(k)\:=\:\nu_0 \epsilon^{1/3} k^{-1} k_0^{-1/3} $ and the above integral needs to be cut-off at a lower limit $ k_0 $ leading to $ V_0^2\:=\:\frac{A}{8\pi}\:\frac{\epsilon^{2/3}}{k_0^{2/3}} $. It is this $ V_0 $ which sets the scale for the velocity in the boundary layer. We now need to specify the region of z-space in which we are interested. As already discussed, the law of the wall in eqn.(1) holds somewhat away from the boundary. The relevant length scale is $ k_D^{-1}\:=\:\frac{\nu(k_0)}{V_0}\:=\:\sqrt{8\pi} \nu_0^{3/2} \frac{k_0^{-1}}{A^{1/2}} $. Our interest is in the value of $ z $ for which
$ k_D z > 1 $ and from eqn.(10), we find the correction to $ V_0^2 $ is 
\bea
<u_1 u_3>\:=\:-V_0^2 [1-\frac{5}{6} \frac{k_D}{k_0} \frac{1}{z_{+}}]
\eea
where $ z_{+}=k_D z $ (Note that for small $ k_0 $ we can set $ e^{-k_0 z}\:\sim\:1 $). The quantity $ A/{\nu^3} $ is a universal number [10-12] known from the study of isotropic homogeneous turbulence. Using that universal value we find
\bea
<u_1 u_3>\:=\:-V_0^2[1-\frac{5}{6} \sqrt{\frac{5\pi}{3}} \frac{1}{z_{+}}]
\eea
We now return to eqn.(3) and observe that the constant can be calculated by going far away from the wall, where the viscous stress is negligible. This leads to
\bea
\nu \frac{\partial V_1}{\partial z}\:=\:<u_1 u_3>\:+\:V_0^2
\eea
(A more careful calculation would yield a correction term in the above equation which vanishes for large Reynold's number). Using eqn.(13) for the fluctuating stress tensor and integrating, we get
\bea
\frac{V_x}{V_0}\:=\:\frac{5}{6}\sqrt{\frac{5\pi}{3}}\:ln z_{+}\:+\:constant
\eea
leading to
\bea
\kappa\:=\:\frac{6}{5}\sqrt{\frac{3}{5\pi}}\:\simeq\:0.52
\eea
The experimental values of $ \kappa $ cluster around 0.42.
\par
In summary, we have used a randomly forced Navier Stokes' equation to calculate the fluctuating stress tensor and have thereby obtained the logarithmic law of the wall. The strong point of the procedure is the ability to calculate the universal amplitude appearing in the velocity profile. Our calculated number is in reasonably good accord with the experimentally measured value.
\section{Acknowledgement}
One of the authors (AKC) sincerely acknowledges partial financial support from C. S. I. R., India.

\end{document}